\documentclass[twocolumn,epjc3]{svjour3}          

\RequirePackage[numbers,sort&compress]{natbib}
\RequirePackage[colorlinks,citecolor=blue,urlcolor=blue,linkcolor=blue]{hyperref}

\usepackage{listings}
\pdfoutput=1
\usepackage[utf8]{inputenc}
\usepackage[english]{babel}
\usepackage{hyperref}

\usepackage{lmodern}
\usepackage[safe]{textcomp}
\usepackage{bbm}
\usepackage{amstext}
\usepackage{amsfonts}
\usepackage{graphicx}
\usepackage{amsmath}
\usepackage{amssymb}
\usepackage{amsfonts}
\usepackage{mathtools}

\usepackage{array}                
\usepackage{xcolor} 
\usepackage{slashed}
\usepackage[absolute]{textpos}
\usepackage{multirow}

\usepackage{relsize}

\journalname{Eur. Phys. J. C}

\usepackage{xcolor}
\definecolor{darkblue}{rgb}{0,0,0.7}

\newcommand{\ie}{i.e.\ }
\newcommand{\eg}{e.g.\ }

\newcommand{\wrt}{w.r.t.\ }

\newcommand{\bit}{\begin{itemize}}
\newcommand{\eit}{\end{itemize}}
\newcommand{\bce}{\begin{center}}
\newcommand{\ece}{\end{center}}
\newcommand{\bea}{\begin{eqnarray}}
\newcommand{\eea}{\end{eqnarray}}
\newcommand{\be}{\begin{equation}}
\newcommand{\ee}{\end{equation}}
\newcommand{\ba}{\begin{align}}
\newcommand{\ea}{\end{align}}
\newcommand{\beas}{\begin{eqnarray*}}
\newcommand{\eeas}{\end{eqnarray*}}
\newcommand{\bes}{\begin{equation*}}
\newcommand{\ees}{\end{equation*}}
\newcommand{\bas}{\begin{align*}}
\newcommand{\eas}{\end{align*}}

\newcommand{\as}{\alpha_{s}}
\newcommand{\lb}{\left(}
\newcommand{\rb}{\right)}

\newcommand{\calM}{\mathcal{M}}

\newcommand{\calO}{\mathcal{O}}

\newcommand{\rd}{\mathrm d}

\newcommand{\TeV}{\mathrm{TeV}}
\newcommand{\GeV}{\mathrm{GeV}}
\newcommand{\MeV}{\mathrm{MeV}}

\newcommand{\fb}{\mathrm{fb}}

\newcommand{\LO}{\mathrm{LO}}
\newcommand{\LOgg}{\mathrm{LO(gg)}}
\newcommand{\LOqg}{\mathrm{LO(qg)}}
\newcommand{\LOqq}{\mathrm{LO(q\bar{q})}}
\newcommand{\NLO}{\mathrm{NLO}}
\newcommand{\NLOsv}{\mathrm{NLOsv}}

\newcommand{\NNLOsv}{\mathrm{NNLOsv}}
\newcommand{\NNLOsvp}{\mathrm{NNLOsv\textquotesingle}}

\newcommand{\SM}{\mathrm{SM}}

\definecolor{bluemar}{rgb}{0,0,.5}
\definecolor{redmar}{rgb}{.8,0,0}
\definecolor{greenmar}{rgb}{0,.5,0}

\newcommand{\gam}{\gamma}
\newcommand{\ep}{\epsilon}
\newcommand{\ff}[2]{\displaystyle\frac{#1}{#2}}
\newcommand{\logs}{\log\ff{{\mu^2}}{m_{\gam\gam}^2}}
\newcommand{\logsq}{\log^2\ff{{\mu^2}}{m_{\gam\gam}^2}}

\renewcommand{\Re}{\mathrm{Re}}
\renewcommand{\Im}{\mathrm{Im}}
\newcommand{\Mbkg}{\calM_{\mathrm{bkg}}}
\newcommand{\Msig}{\calM_{\mathrm{sig}}}
\newcommand{\DD}[2]{\mathcal{D}_{#2}(#1)}
\newcommand{\MSbar}{\overline{\mathrm{MS}}}
\newcommand{\asontwopi}{\left(\ff{\as(\mu)}{2\pi}\right)}

\graphicspath{{./figures/}}
\allowdisplaybreaks

\usepackage{tocloft}
\makeatletter
\renewcommand{\numberline}[1]{%
  \@cftbsnum #1\@cftasnum~\@cftasnumb%
}
\newcommand{\eqnum}[1]{\leavevmode\hfill\refstepcounter{equation}\label{#1}\textup{\tagform@{\theequation}}}
\makeatother
 
\begin{document}
\title{Signal-background interference effects in Higgs-mediated diphoton production beyond NLO}

\author{
Piotr~Bargie\l{}a\thanksref{e1,ox}\and 
Federico~Buccioni\thanksref{e2,ox,tum}\and
Fabrizio~Caola\thanksref{e3,ox}\and
Federica~Devoto\thanksref{e4,ox}\and
Andreas~von~Manteuffel,\thanksref{e5,mich}\and
Lorenzo~Tancredi,\thanksref{e6,tum}
}

\thankstext{e1}{e-mail: piotr.bargiela@physics.ox.ac.uk}
\thankstext{e2}{e-mail: federico.buccioni@tum.de}
\thankstext{e3}{e-mail: fabrizio.caola@physics.ox.ac.uk}
\thankstext{e4}{e-mail: federica.devoto@physics.ox.ac.uk}
\thankstext{e5}{e-mail: vmante@msu.edu}
\thankstext{e6}{e-mail: lorenzo.tancredi@tum.de}

\institute{\label{ox}
Rudolf Peierls Centre for Theoretical Physics, University of 
Oxford,
Clarendon Laboratory, Parks Road, Oxford OX1 3PU, UK
\and
{\label{tum}}
Physics Department, Technical University of Munich, James-Franck-Straße 1, 85748 Garching,
Germany
\and
{\label{mich}}
Department of Physics and Astronomy, Michigan State University,
East Lansing, Michigan 48824, USA 
}

%
\date{}
%

\maketitle

\begin{abstract}
In this paper we consider signal-background interference effects in
Higgs-mediated diphoton production at the LHC. After reviewing earlier
works that show how to use these effects to constrain the Higgs
boson total decay width, we provide predictions beyond NLO accuracy for
the interference and related observables, and study the impact of QCD
radiative corrections on the Higgs width determination. 
In particular, we use the so-called soft-virtual approximation to
estimate interference effects at NNLO in QCD. The inclusion of these effects
reduce the NNLO prediction for the total Higgs cross-section 
in the diphoton channel by about 1.7\%.
We study in detail the impact of QCD corrections on the Higgs-boson line-shape
and its implications for the Higgs width extraction.
Assuming an experimental resolution of about 150~$\MeV$ on interference-induced
modifications of the Higgs-boson line-shape, 
our NNLO analysis shows that one could constrain the Higgs-boson total width
to about 10-20 times its Standard Model value. 
\end{abstract} 
\vfill
\setcounter{tocdepth}{2}
\tableofcontents

\section{Introduction} \label{intro}
Only a few months ago we celebrated the tenth-year anniversary of the
Higgs boson discovery at the CERN Large Hadron Collider
(LHC)~\cite{ATLAS:2012yve,CMS:2012qbp}.  Since then, enormous efforts
both from the theory and the experimental communities have been
devoted to a precise determination of the Higgs boson properties such
as its mass,  decay width and  couplings to other Standard Model
(SM) particles.  Indeed,
an in-depth exploration of the Higgs sector is one of
the main goals of the current and future LHC
precision physics program~\cite{Cepeda:2019klc}.

The predominant mechanism of Higgs boson 
production at the LHC is gluon fusion.
The $H\to\gamma\gamma$ and $H\to ZZ^*\to 4l$ decay channels, despite
having small branching ratios, provide a very clean environment
for the study of Higgs-boson properties.
%
Measurements in the diphoton channel allowed for a determination of the
Higgs boson mass 
with an uncertainty of $260~\MeV$, roughly half of which is
systematic~\cite{CMS:2020xrn}.
The $ZZ$ channel allows for an even better determination, thanks to a
very good experimental control on the final-state leptons. Indeed, in
this channel the Higgs boson mass has been measured with an accuracy
of $184~\MeV$~\cite{ATLAS:2022net}. In this case, the uncertainty
is mostly dominated by statistics, which accounts for $180~\MeV$.

Measuring the Higgs boson total decay width $\Gamma_H$ is much more
challenging, because of the extremely narrow nature of the Higgs resonance.
Indeed, the predicted SM value of roughly $4~\MeV$ has to be
confronted with an experimental sensitivity of the order of
$1-2~\GeV$~\cite{CMS:2017dib,ATLAS:2022tnm}. Therefore, one has to
resort to indirect analysis to extract bounds on the Higgs boson
width. One option is to perform a global fit of SM parameters,
e.g. within the context of the Standard Model Effective theory
~\cite{Brivio:2019myy,Ellis:2020unq,Ethier:2021bye}.  However, it has
also been pointed out in the literature that one can harness the
sensitivity of specific observables on the Higgs width to constrain
the latter.  One proposal is to exploit the peculiarities of the
off-shell Higgs cross-section in the four-lepton
channel~\cite{Kauer:2012hd,Caola:2013yja,Campbell:2013una}, which
allows one to probe values of $\Gamma_H$ as small as the SM
one~\cite{CMS:2022ley,ATLAS:2018jym}. However, such a technique relies
on some underlying theoretical assumptions, see
e.g. Refs~\cite{Englert:2014aca,Logan:2014ppa,Azatov:2022kbs}.\footnote{The
  model-dependence of this approach can be alleviated by combining
  results in the gluon fusion and VBF
  channels~\cite{Campbell:2015vwa}.} Another proposal is to exploit
signal-background interference effects in the diphoton
channel~\cite{Martin:2012xc,Dixon:2013haa}. Such effects are expected
to shift the Higgs invariant mass distribution peak by a value that
depends on the Higgs boson width. In the SM, the shift turns out to be
quite small. Initial theoretical analyses estimated it to be of about
$50-100~\MeV$~\cite{Dixon:2013haa,deFlorian:2013psa}. A more robust
estimate, that took into account realistic experimental conditions, was
conducted by the ATLAS collaboration and found a somewhat
smaller mass-shift of about $40~\MeV$~\cite{ATLAS:2016kvj}.
The predicted
experimental sensitivity on the mass-shift at the LHC is of about few
hundred $\MeV$~\cite{ATLAS:2022tnm,CMS:2022sfn}, which translates to
an upper bound on the Higgs width of about $5-30$ times the standard
model
value~\cite{Dixon:2013haa,deFlorian:2013psa,Cieri:2017kpq,Campbell:2017rke}.
While such bounds are not as constraining as the ones obtained from
the off-shell method, they do not suffer from the same model
dependence and provide therefore important complementary
information.

For a reliable extraction of the Higgs boson width from the
mass-shift, one needs a good theoretical control on the latter.
The preliminary LO studies of Ref.~\cite{Martin:2012xc},
which considered the dominant $gg\to\gamma\gamma$ channel, showed
that the apparent mass-shift could be of $\calO(100-200\,\MeV)$ for
typical collider energies and setup.
This LO analysis was later refined in Ref.~\cite{deFlorian:2013psa} via 
the inclusion of $qg$, $\bar{q}g$ and $q\bar{q}$ initiated processes,
which account for a shift of $\calO(30\,\MeV)$, carrying an opposite
sign with respect to the $gg$ case.
The impact of higher-order QCD correction was soon after addressed 
in Ref.~\cite{Dixon:2013haa}, which also explicitly noted
that a measurement of the mass-shift
could be used to put indirect bounds on $\Gamma_H$. The results of
Ref.~\cite{Dixon:2013haa} were later confirmed in
Ref.~\cite{Campbell:2017rke}. Additionally, the impact of small-$p_T$
resummation~\cite{Cieri:2017kpq} and extra hard QCD
radiation~\cite{Coradeschi:2015tna} on the mass-shift were also
studied.
One of the main outcomes of Ref.~\cite{Dixon:2013haa}, is that NLO QCD
corrections account for a large $\calO(40\%)$ effect on the mass
shift, hence highlighting the relevance of higher-order corrections.
Furthermore, the bulk of the effect comes from the low
$p_{T,\gamma\gamma}$
region~\cite{Martin:2013ula,Dixon:2013haa,Coradeschi:2015tna}.

The primary goal of this article is to improve on 
the current predictions for the mass-shift and extend the 
analysis beyond NLO QCD. For a long time, this was prevented by the
lack of the relevant multi-loop amplitudes for the $gg\to\gamma\gamma$
continuum production. This bottleneck has recently been overcome
and analytic results for the three-loop helicity amplitudes for
$gg\to\gamma\gamma$~\cite{Bargiela:2021wuy}, as well as for the two-loop
ones for $\gamma\gamma+j$
production~\cite{Badger:2021imn,Agarwal:2021vdh} are now available.  
In principle, this -- together with the well-known
analogous results for the $gg\to H\to \gamma\gamma$ signal -- allows
for a complete NNLO evaluation of the signal-background interference
and hence of the mass-shift. In practice however, such an endeavour is
non trivial as it requires very good numerical control of the two-loop 5-point
scattering amplitudes in soft-collinear regions. In this paper, we perform a
first step towards the full NNLO calculation and work in the so-called
\emph{soft-virtual} approximation. Within such approximation, one retains
the full information of virtual corrections and soft real
emissions, but neglects the impact of hard radiation. Since the bulk
of the interference is dominated by the region where the
$\gamma\gamma$ pair has a low transverse
momentum~\cite{Martin:2013ula,Dixon:2013haa,Coradeschi:2015tna}, we
expect this approximation to work quite well in this case.

The remaining of this paper is organised as follows.
In Section~\ref{se:thbkg} 
we review the theoretical background of Higgs interferometry in diphoton
production. 
In Section~\ref{se:highordcorr} we provide details of our calculation,
discussing all  necessary ingredients with a special focus on
the soft-virtual approximation for colour-singlet production.
In Section~\ref{se:results} we discuss our phenomenological results. First, we validate
the soft-virtual approximation at NLO, and then use it to estimate
the impact of NNLO QCD corrections. We finally conclude in Section~\ref{se:conclusions}.


\section{Theoretical background} \label{se:thbkg}
In this section, we briefly review the main aspects of
signal-background interference for Higgs-mediated diphoton production
at the LHC. For the sake of illustration, we discuss the main features
of the interference at LO, focusing on the gluon-fusion channel. The complete
analysis will be presented in Section~\ref{se:highordcorr}.
\subsection{Higgs interferometry}
\label{se:higgsint}
We consider diphoton production at the LHC in the gluon-fusion
channel. At order $\as^2$ two main mechanisms contribute: the
Higgs-mediated process $gg\to H\to \gamma\gamma$ and the continuum
process $gg\to\gamma\gamma$. We refer to the former as our ``signal''
and to the latter as our ``background''. Schematically, we write the
scattering amplitude for this process as
\begin{equation}
\label{eq:sigbkgamp}
  \calM_{gg\to\gamma\gamma} = \frac{\Msig}{m_{\gamma\gamma}^2 - m_H^2 + i \Gamma_H m_H}
  + \Mbkg,
\end{equation}
where $m_{\gamma\gamma}$ is the diphoton invariant mass and where we have
explicitly factored out the Higgs-boson propagator. 
In order to improve readability, we dropped helicity labels, 
which are understood.
It is helpful to further separate the real and 
imaginary parts of $\calM_{\mathrm{sig,bkg}}$, \ie
\begin{equation}
\label{eq:reimsplit}
\calM_{\mathrm{sig,bkg}} = \mathrm{Re}\calM_{\mathrm{sig,bkg}} + i\,\mathrm{Im}\calM_{\mathrm{sig,bkg}}\,.
\end{equation}
Since we will be ultimately interested in the
diphoton invariant-mass distribution, 
we need to consider the square of the amplitude in Eq.~\eqref{eq:sigbkgamp},
which reads
\begin{align}
|\calM_{gg\to\gamma\gamma}|^2 &= \frac{|\Msig|^2}{\lb
  m_{\gamma\gamma}^2 - m_H^2\rb^2 + \Gamma_H^2 m_H^2} + |\Mbkg|^2
\nonumber \\ &+ 2\,\mathrm{Re}\lb
\ff{\Msig}{m^2_{\gamma\gamma}-m_H^2+i \Gamma_H
  m_H}\Mbkg^{\dagger}\rb.
\label{eq:sqamp}
\end{align}
The invariant-mass distribution can then be schematically organised as follows
\begin{equation}
\ff{\rd\sigma}{\rd m_{\gamma\gamma}} \sim |S|^2 + |B|^2 + I\,,
\end{equation}
where the three terms $S$, $B$ and $I$ are in one-to-one correspondence
with those on the right-hand side of Eq.~\eqref{eq:sqamp}.
The signal-background interference part is the last one in the equation above.
One can get insight on the structure of the interference contribution by
further separating it into a so-called ``real part'' $I_\Re$ and 
an ``imaginary part'' $I_\Im$~\cite{Dicus:1987fk}, 
i.e. $I = I_\Re + I_\Im$.
These two components can be expressed through the actual real and imaginary
parts of the amplitudes in Eq.~\eqref{eq:reimsplit} and 
are given by
\begin{align}
&I_{\mathrm{Re}}  \propto \ff{2}{(m_{\gamma\gamma}^2-m_H^2)^2+\Gamma_H^2
  m_H^2}\left(m_{\gamma\gamma}^2-m_H^2\right) \times  \nonumber \\
&\times \left[\Re\Mbkg \Re\Msig + \Im\Mbkg \Im\Msig\right],  \\ \nonumber
&I_{\mathrm{Im}}  \propto \ff{2}{(m_{\gamma\gamma}^2-m_H^2)^2+\Gamma_H^2
  m_H^2} \Gamma_H m_H \times \\ 
  &\times \left[\Re\Mbkg \Im\Msig - \Im\Mbkg \Re\Msig\right].
  \label{eq:realim}
\end{align}
\begin{figure}[t!]
  \begin{center}
  \includegraphics[scale=0.34,trim=123 50 88 50]{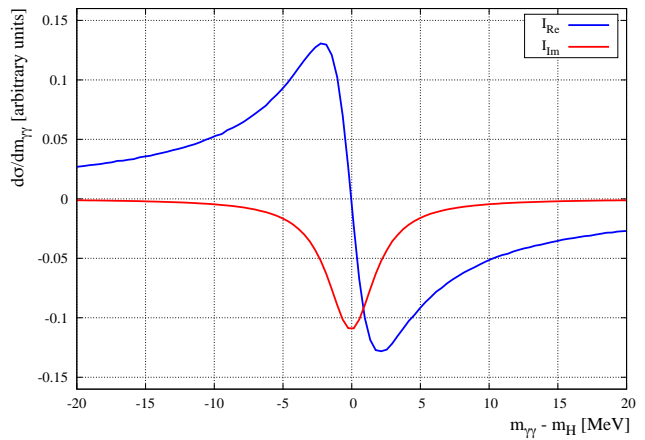}
  \caption{Real and imaginary parts of the signal-background
    interference terms, see text for details. This figure is just for illustration
  purposes. Our best prediction for these curves will be described
  in details in Sec.~\ref{se:results}.}
  \label{fig:intreim}
  \end{center}
\end{figure}
It is clear from these equations that the real part of the interference is an antisymmetric function of the diphoton invariant mass $m_{\gamma \gamma}^2$ 
around the Higgs resonance, and therefore does not contribute to the
total cross-section. 
This is not the case for the imaginary part, which is instead
symmetric around the resonance. This is illustrated in Fig.~\ref{fig:intreim}, where
we plot $I_\Re$ and $I_\Im$ in the $gg$ channel up to NLO, to better visualise
their independent effects.

Naively, one may think that the relative impact of the interference on the
Higgs total cross section could be quite sizeable, since the signal is a
two-loop process while the background starts at one loop. Because of
this, one may expect a loop-enhancement factor of the interference
with respect to the signal. However, a close inspection of
Eq.~\eqref{eq:realim} shows that the
contribution to $I_{\Im}$ from the imaginary part of the background
is strongly suppressed at
leading order.  This follows from the fact that the Higgs boson, being
a scalar, only decays into a pair of photons with identical
helicity. 
In turn, if the photons have equal helicities 
the imaginary part of the background at leading order
vanishes, unless the process is mediated by a massive
quark. In our calculation we keep full dependence on the bottom quark
at leading order but, since its contribution is mass-suppressed, 
the net effect is small and can be safely neglected at higher orders.
Starting from two loops, the relevant background helicity amplitudes 
develop an imaginary part. This leads to $I_\Im$ having a destructive impact of around
$1$-$2\%$~\cite{Dixon:2003yb,Campbell:2017rke}. As far as the real
part is concerned, there is no mass suppression at LO. Although such
effect does not contribute to the total cross section,
c.f. Eq.~\eqref{eq:realim}, it leads to a non-negligible shift of
events around the diphoton invariant-mass peak, as it was first noted
in Ref.~\cite{Martin:2012xc}.

Moreover, it was observed in Ref.~\cite{Dixon:2013haa} that the integrated interference $I$
has a different dependence on the Higgs boson production and decay
couplings compared to the integrated signal. Specifically, if we
schematically denote with $\lambda_g$ and $\lambda_{\gamma}$ the Higgs
couplings to gluons and photons, we find that the signal is
proportional to $\lambda_g^2 \lambda_{\gamma} ^2 / \Gamma_H$, while
the interference is only proportional to $\lambda_g \lambda_{\gamma}$.
The main idea of Ref.~\cite{Dixon:2013haa} is to exploit this
intertwined dependence on the Higgs boson
couplings and its decay rate to extract information on $\Gamma_H$
and to look for possible deviations from its SM value.
Current experimental measurements constrain the Higgs-diphoton rate
to be the same of its SM prediction to within 10\%~\cite{ATLAS:2022qef,ATLAS:2022fnp,CMS:2022wpo}. At this level of
accuracy, one can neglect the small impact of the destructive interference
$I_\Im$  on the cross section. One then schematically writes this
experimental observation as
\begin{equation}
  \ff{\lambda_g^2 \lambda_{\gamma} ^2}{\Gamma_H} \approx
  \ff{\lambda_{g,\rm SM}^2 \lambda_{\gamma,\rm SM} ^2}{\Gamma_{H,\rm SM}},
  \label{eq:flat}
\end{equation}
where $\lambda_{i}$, $\Gamma_H$ are the ``true'' Higgs-boson couplings and
width and $\lambda_{i,\SM}$, $\Gamma_{H,\SM}$ are their respective SM
predictions. Eq.~\eqref{eq:flat} holds in particular under appropriate rescalings
of $\lambda_{i,\SM}$ and $\Gamma_{H,\SM}$, 
\eg $\lambda_i \to \kappa \lambda_{i,\SM}$ and
$\Gamma_H \to \kappa^4 \Gamma_{H,\SM}$,
so it does not allow for a simultaneous direct extraction of couplings and width.
However, if one supplements Eq.~\eqref{eq:flat}
with the observation that $I_\Re\propto \lambda_g\lambda_\gamma$,
then one obtains
\begin{equation}
  \ff{I_\Re}{I_{\Re,\SM}} \propto \sqrt{\ff{\Gamma_H}{\Gamma_{H,\SM}}}.
\end{equation}
Hence, a measurement of $I_\Re$ allows for an extraction of $\Gamma_H$,
provided that the SM prediction $I_{\Re,\SM}$ is under good theoretical
control.

As mentioned before, the main effect of $I_\Re$ is to distort
the Higgs invariant-mass distribution,
effectively shifting the position of the peak. This translates
into an apparent mass-shift with respect to the SM value~\cite{Martin:2012xc}.
A measurement of such mass-shift can then be used to constrain $\Gamma_H$.
In the next section, we review how one can extract the mass-shift from
the knowledge of the Higgs invariant-mass distribution.

\subsection{Extraction of the mass-shift}
\label{se:mass_shift}
 One first
remark is that the Higgs resonance is extremely narrow, and therefore
strongly smeared by the finite detector resolution.  To properly take
this into account, one should convolve theoretical prediction with
a full detector simulation. However, such a study can only be carried
out by experimental collaborations and it is outside the scope of this
theoretical work. In this paper we estimate the mass-shift using two
different, yet related, proxies for it that have been presented in the
literature.  Namely, we will consider the first
moment of the diphoton invariant-mass distribution~\cite{Martin:2012xc} and
a full gaussian fit to it~\cite{Dixon:2013haa}.

Before reviewing these techniques, we stress that they are inherently
different, so one should not expect identical results for the mass shift.
In other words, we can think of these procedures as a measurement of
an observable that is strongly correlated, yet not identical, to the mass
shift that experimentalists would measure. Nevertheless, we may imagine
that they capture similar physics, and hence they should receive comparable
radiative corrections. In Section~\ref{se:results}, we will see that this
is indeed the case. This gives us confidence that, even if our theoretical
predictions for the absolute values of the mass shift are inherently
limited by our experimental modeling, our results for the QCD $K$-factors
are quite robust. 

We now describe the two methods in some detail. In both cases, we
simulate detector effects by smearing the diphoton invariant-mass
distribution using a Gaussian function with
$\sigma=1.7~\GeV$~\cite{ATLAS:2016kvj}.

The first-moment method~\cite{Martin:2012xc} is based on the observation
that, from the theoretical side, a very simple way to access the mass shift is
to consider the first moment of the invariant-mass
distribution, i.e.
\begin{equation}
\label{eq:firstmom}
\langle m_{\gam \gam}\rangle = \ff{1}{\sigma_{gg \to \gam \gam}}\int d
m_{\gam \gam} m_{\gam \gam} \ff{d \sigma_{gg \to \gam \gam}}{dm_{\gam
    \gam}}
\end{equation}
where $\sigma_{gg \to \gam \gam}$ is the fiducial cross-section.  The
mass shift is then defined as \be \Delta m_{\gam \gam} = \langle
m_{\gam \gam}\rangle_{\mathrm{sig+int}}-\langle m_{\gam
  \gam}\rangle_{\mathrm{sig}}.  \ee The main advantage of this method
is that it is theoretically very clean.  Also, it is not very
sensitive to overall normalisation issues, but rather focuses on the
position of the peak.  On the practical level however, it requires 
exquisite resolution on the invariant mass distribution which is very
hard to achieve experimentally.  It also strongly depends on
the technical details of the theoretical analysis.
For example, in Ref.~\cite{Martin:2012xc}
the author found that the mass-shift strongly depends on the choice of
the upper and lower integration boundaries in Eq.~\eqref{eq:firstmom}.
Indeed, a change of $\calO(1\,\mathrm{GeV})$ in such
choice modifies the mass-shift estimate by almost $20\%$ at leading
order. 

A different proposal that addresses at least some of the shortcomings
of the first-moment method is to simply perform a Gaussian fit of the
diphoton invariant mass distribution~\cite{Dixon:2013haa}. One then
extracts the mass shift by comparing predictions obtained with and
without including interference effects. It was argued in
Ref.~\cite{Dixon:2013haa} that this method is more resilient against
specific details of analysis with respect to the first-moment one.

These two methods predict a mass-shift of $\mathcal O(100)~\MeV$ at
LO.  More precisely, the first-moment technique gives results in the
interval $\Delta m_{\gamma\gamma} \in \{-250,-150\}~\MeV$ depending on
the integration window in
Eq.~\eqref{eq:firstmom}~\cite{Martin:2012xc}. The likelihood fit of
Ref.~\cite{Dixon:2013haa} instead predicts $\Delta m_{\gamma\gamma} =
-120~\MeV$. As we have stressed before, these two methods measure
correlated yet slightly different observables, so one should not
expect identical results. 

\section{Higher-order QCD corrections} \label{se:highordcorr}
\begin{figure*}[t!]
  \begin{center}
  \includegraphics[scale=0.8]{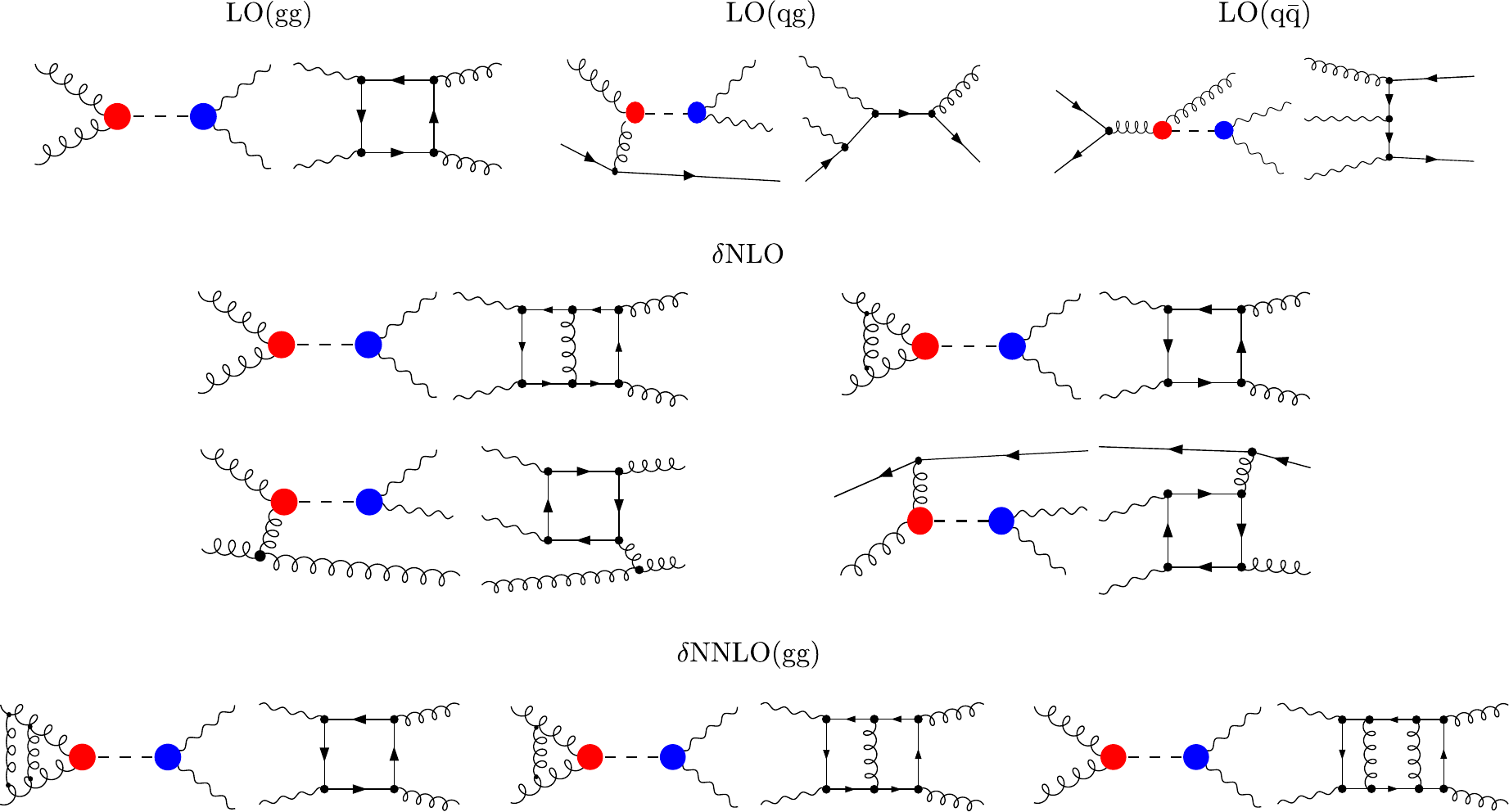}
  \vspace{7mm}
  \caption{Representative diagrams contributing to the interference up to order $\alpha_s^4$. The red dot denotes
  the effective vertex described through the Lagrangian in
  Eq.~\eqref{eq:eff_lag}, while the blue dot denotes the Higgs decay
  via heavy quark loops and W boson, described in Eq.~\eqref{eq:Haadecay}.}
  \label{fig:diagrams}
  \end{center}
\end{figure*}

In this section, we present the technical details of our calculation.
We start by discussing LO predictions for the interference. At order
$\as^2$, we need to consider three different partonic channels, see
Fig.~\ref{fig:diagrams}~(top). We then write
\begin{equation}
  \LO = \LOgg + \LOqg + \LOqq\,.
  \label{eq:loch}
\end{equation}
At this order, we compute the signal retaining the full top- and
bottom- mass dependence. For the background, we neglect top-mediated
contributions, since they are heavily suppressed. As we have explained
in Section~\ref{se:higgsint}, the bottom-quark contributions are the only
ones leading to a non-vanishing $I_\Im$ at this order. 
This effect is very small~\cite{Campbell:2017rke}, so we neglect it
at higher orders. Because of this, when computing (N)NLO corrections we
 consider Higgs production in the heavy-top effective theory, described
by the following Lagrangian
\begin{equation}
  \label{eq:eff_lag}
\mathcal{L}_{\mathrm{eff}} = -\frac{\lambda}{4} \,H G_{\mu\nu}^a G^{\mu \nu}_a,
\end{equation}
where $\lambda = -\as/(3\pi v) + \calO(\as^2)$ is the bare coupling and $v$ is the Higgs
vacuum expectation value.\footnote{We report higher-order corrections
  to $\lambda$ in~\ref{app:amp_def}, see Eq.~\eqref{eq:lambdahighord}.} At higher-orders,
we still treat the Higgs decay to photons at LO, with exact mass
dependence. Indeed, QCD radiative corrections are known to be
small~\cite{Djouadi:1990aj}. For convenience, we report the relevant
amplitude in~\ref{app:amp_def}.
As far as the background is concerned,
we set the bottom mass to zero beyond LO.\footnote{Let us stress 
once again that the main reason for retaining the exact bottom-mass 
dependence at LO is
to generate an imaginary part that would not be present otherwise. Beyond
one-loop, the massless amplitude also develops an imaginary part so
bottom-mass effects are subleading and can be safely discarded.}

It is well-known that the $gg$ channel is the dominant one for the signal
process.
For the interference, this statement is somehow weakened but still true.
Indeed, the $qg$ channel (see Fig.~\ref{fig:diagrams}) accounts for about
30\% of the result~\cite{deFlorian:2013psa}. While it is
easy with current technology to perform a full NLO analysis,
in this paper we are mostly interested in the impact of NNLO corrections.
We therefore focus our attention on the $gg$ channel only.

%
%

Strictly speaking, if we include only gluon-induced processes, NLO
corrections are given by only the first three out of the four $\delta \NLO$
diagrams of Fig.~\ref{fig:diagrams}. 
Of course, this would imply that one should evolve parton
distribution functions (PDFs) in the same approximation. However, one
can also retain formally subleading effects. In particular, the
quark-induced contribution in the third row of
Fig.~\ref{fig:diagrams} is linked to the $gg$ channel through PDF
evolution. One may then expect that its inclusion in the DGLAP evolution
of the PDFs would alleviate the
factorisation-scale dependence of the result. 
Therefore, following Ref.~\cite{Dixon:2013haa}, in our
analysis we use standard parton distributions and include the last
diagram in the third row of Figure.~\ref{fig:diagrams}. 
In what follows, we call `$\delta$NLO'
the contribution coming from the second and third rows of that figure. We
stress that this is \emph{not} the full NLO correction to the
interference. It may also be instructive to consider only gluon-induced
diagrams at this order. Indeed, comparing this against our default setup may give us
a handle on the large-gluon approximation. In what follows, we define
corrections computed using only the first three diagrams of the 
block of diagrams ($\delta$NLO) in
Fig.~\ref{fig:diagrams} as `$\delta$NLO(gg)'. Correspondingly we call the
the full NLO result
\begin{equation}
  \NLO = \LO + \delta \NLO,~~ \NLO(gg) = \LO + \delta\NLO(gg),
\end{equation}
with $\LO$ defined in Eq.~\eqref{eq:loch}.

We now move to NNLO corrections. As we have said, at this order we
work in the soft-virtual approximation. This is justified by
the fact that, at least at lower orders, interference effects are
stronger in the region where the diphoton transverse momentum is
small~\cite{Martin:2013ula,Dixon:2013haa}. The soft-virtual
approximation and various refinements of it have been extensively
adopted for Higgs
predictions~\cite{Catani:2001ic,Moch:2005ky,Ball:2013bra,deFlorian:2014vta,
  Anastasiou:2014vaa}. The
main advantage of working in this limit is that the
only process-dependent part is encoded in purely virtual
contributions, see e.g. Ref.~\cite{deFlorian:2012za} for an explicit
derivation.

Here we only sketch the structure of the soft-virtual approximation, and
refer the reader to Ref.~\cite{deFlorian:2012za} for more details. We write
the fully-differential hadronic cross section for diphoton production in the
$gg$ channel as
\begin{align}
  d\sigma(\tau,y,\{\theta_i\}) = \int d\xi_1 d\xi_2 dz f_g(\xi_1,\mu_F)
  f_g(\xi_2,\mu_F)\nonumber 
  \\
  \times\,\delta(\tau-\xi_1\xi_2 z) d\hat \sigma
  \left(z,\hat y,\{\hat\theta_i\},\as,\ff{Q^2}{\mu_R^2},\ff{Q^2}{\mu_F^2}\right).
  \label{eq:dsigma}
\end{align}
In this equation, $\tau = Q^2/s_{\rm had}$ with $Q$ being the
invariant mass of the diphoton system and $s_{\rm had}$ the square of
the collider energy, $y$ is the rapidity of the diphoton system in the
laboratory frame and $\theta_{i}$ are a set of variable that fully
describe the diphoton system (e.g. scattering angles). 
Variables with hats represent the corresponding partonic
quantities. Eq.~\eqref{eq:dsigma} is fully differential in
the kinematics of the diphoton system, but retain no
information on extra QCD radiation. We note that the partonic
variables $\{\hat \theta_i\}$ depend on their hadronic counterparts
$\{\theta_i\}$ and also on $z,\xi_1,\xi_2$. Finally, $\as =
\alpha_s(\mu_R)$ is the renormalised QCD coupling constant evaluated at scale $\mu_R$. 

Major simplifications occur in the soft $z\to1$ limit. First, in this
case the rapidity dependence of Eq.~\eqref{eq:dsigma} is entirely
fixed by the inclusive cross section, up to power
corrections~\cite{Bolzoni:2006ky,Becher:2007ty,Bonvini:2010tp}. The
partonic cross section in Eq.~\eqref{eq:dsigma}  simplifies to
\begin{align}
&d\hat \sigma \left(z,\hat
y,\{\hat\theta_i\},\as,\ff{Q^2}{\mu_R^2},\ff{Q^2}{\mu_F^2}\right)\approx
\nonumber\\
&d\hat\sigma_{\rm Born}(\{\hat\theta_i\},\as) \,z\, G\left(
z,\as,\ff{Q^2}{\mu_R^2},\ff{Q^2}{\mu_F^2} \right),
\label{eq:dsigma_sv}
\end{align}
where we  neglected power corrections in $(1-z)$. In
Eq.~\eqref{eq:dsigma_sv}, $d\sigma_{\rm
  Born}(\{\hat\theta_i\},\as)\delta(1-z)$ is the Born cross section
and $G$ is the inclusive coefficient function in the soft limit,
normalised such that $G = \delta(1-z) + \mathcal O(\as)$. In the soft
limit, the diphoton kinematics is identical to the LO one, except that the
partonic center-of-mass energy is rescaled by $z$. Such a rescaling
factor can be absorbed by boosting an individual leg, i.e. by using
the partonic momenta $\xi_{1,\rm Born} \to z \,\xi_{1,\rm Born}, 
\xi_{2,\rm Born}\to \xi_{2,\rm Born}$ (and vice versa).  Alternatively,
one can boost both legs at the same time, i.e. $\xi_{i,\rm Born} \to
\sqrt{z} \,\xi_{i,\rm Born}$, $i=1,2$. In the soft-limit, the two are
formally equivalent and, in principle, one can consider both and treat their
difference as an uncertainty. In practice, we expect this difference to
be  small~\cite{Bonvini:2013jha}, so for simplicity we always boost
only one leg in this work. 

If we write the perturbative expansion of the coefficient function as
\begin{equation}
\label{eq:Geq}
G\left(z,\as\right) = \delta(1-z) + 
\sum_{n=1}^{\infty}\left(\ff{\as}{2\pi}\right)^n G^{\left(n\right)}(z),
\end{equation}
then in the soft-virtual approximation the individual coefficients $G^{(n)}$
have the form
\begin{equation}
  G^{(n)}\left(z,
  \ff{Q^2}{\mu_R^2},\ff{Q^2}{\mu_F^2}\right) =
    c_0 \delta(1-z)
      + \sum_{k=1}^{2n-1} c_k \mathcal D_k(z),
  \label{eq:g_sv}
\end{equation}
where $c_k = c_k (Q^2/\mu_R^2,Q^2/\mu_F^2)$ and 
$\mathcal D_k$ are the standard plus distributions
\begin{equation}
  \mathcal D_k(z) = \left[
    \frac{\ln^k(1-z)}{1-z}
    \right]_+.
\end{equation}
We present explicit formulas for the NLO and NNLO coefficients,
retaining full scale dependence, in \ref{app:NNLOsv}. 
The coefficients $c_k$ in Eq.~\eqref{eq:g_sv} in principle
depend on the process under consideration. However, it turns out that the only process dependence arises from finite remainders of purely virtual contributions and it is therefore encoded in the coefficient $c_0$. As a consequence, the coefficients of the plus distributions are completely process-independent.

For the analysis in the next section, we also require the signal
process at NNLO accuracy. In principle, computing this exactly does
not pose significant challenges. Nevertheless, we expect the exact result to be very well described by a (refined)
version of the soft-virtual
approximation~\cite{Catani:2001ic,Moch:2005ky,Ball:2013bra,deFlorian:2014vta}.
While the pure soft-virtual prediction provides a good approximation to the interference, this is not the case for the signal.
We therefore employ a
modified soft-virtual approximation to describe the latter.  
Specifically, we follow the approach of Ref.~\cite{Ball:2013bra} which
is known to reproduce the exact NNLO prediction to within few-percent
accuracy. In practice, this amounts to modifying the plus distributions in
Eqs.~(\ref{eq:NLOsv},\ref{eq:NNLOsv}) according to~\cite{Ball:2013bra}
\begin{equation}
  \mathcal D_i(z) \to \mathcal D_i(z) +
  (2-3z+2z^2)\frac{\ln^i\frac{1-z}{\sqrt{z}}}{1-z} - \frac{\ln^i(1-z)}{1-z}.
  \label{eq:marco}
\end{equation}
This modification captures soft emission at next-to-leading power, as
well as part of the hard collinear emission. Such an approximation was
already used for estimating signal-background interference effects for
the $gg\to VV$ process~\cite{Bonvini:2013jha}. In our case, we have
specifically checked that for on-shell Higgs, within the fiducial
volume used in our analysis, the soft-virtual approximation improved
according to Eq.~\eqref{eq:marco} agrees with the exact NNLO result to within
few percent. This is good enough for the kind of accuracy targeted in our
analysis. In the following section, we will refer to this improved soft-virtual
approximation as ``$\NNLOsvp$''.
The reason why we do not
adopt such an improved approximation for the background is because
it is currently unknown how to properly capture next-to-leading power
soft term in this case. Indeed, a simple modification of emission off
external legs works for the Higgs point-like interaction, but not for
the background amplitude. Fortunately, the interference
seems to be dominated by low-$p_T$ physics so the lack of such an
improved approximation is less problematic than for the signal.

We conclude this section by listing the various ingredients of our
calculation. The LO amplitudes, including the quark-mass effects, are
well known, see e.g.~\cite{Ellis:1996mzs} and references therein.  At
NLO, we took the one-loop amplitudes for the background from
Refs.~\cite{Bargiela:2021wuy,Campbell:2016yrh,Bern:2002jx,Agarwal:2021vdh},
borrowing most of them from
\texttt{MCFM}~\cite{Campbell:1999ah,Boughezal:2016wmq,Campbell:2019dru}.
We note that the two-loop amplitude for $gg\to\gamma\gamma$ was first
computed in Ref.~\cite{Bern:2001df}.  NLO predictions were computed
using FKS subtraction~\cite{Frixione:1995ms}.  Finally, the three-loop
$gg\to\gamma\gamma$ amplitude relevant for NNLO corrections was taken
from Ref.~\cite{Bargiela:2021wuy}.  All the tree- and one-loop results
for the interference were cross-checked against a dedicated version of
\texttt{OpenLoops2}~\cite{Buccioni:2019sur,Cascioli:2011va}.


\section{Results} \label{se:results}
In this section, we discuss our main findings. We start
by specifying the setup we are using. 

We employ the so-called $G_\mu$ input scheme for the electroweak
parameters and we use 
$$G_F = 1.16639\cdot 10^{-5}~\mathrm{GeV}^{-2}\,,$$ 
$$
m_W=80.398~\GeV, \quad  m_Z = 91.1876~\GeV$$ 
that give the electroweak coupling
constant $\alpha = 1/132.338$.  
We set the Higgs mass to $m_H=125~\GeV$.
Finally, we use $m_t = 173.2~\GeV$ and $m_b=4.18~\GeV$ for the
top and bottom mass, respectively. We use the
\texttt{NNPDF31\_nnlo\_as\_0118}~\cite{NNPDF:2017mvq} parton
distribution functions as distributed through the \texttt{LHAPDF}
library~\cite{Buckley:2014ana} and we use \textsc{Hoppet}~\cite{Salam:2008qg} 
for various PDFs manipulations. 
The value of the strong coupling
costant is extracted from the PDF set, with
$\as(m_Z)=0.118$.  For the QCD factorisation and renormalisation
scales, we choose the common value $\mu_F = \mu_R =
m_{\gamma\gamma}/2$, where $m_{\gamma\gamma}$ is the invariant mass of
the diphoton system.  All of the results presented in this section
have been derived for the $13.6~\TeV$ LHC, \ie proton-proton
collisions.  We define the fiducial region by imposing the following
set of cuts on the final-state photons
\begin{align}
\label{eq:fiducialsetup}
\quad\quad & p_{T,\gamma} > 20\,\GeV, \quad\quad
\sqrt{p_{T,\gamma_1} p_{T,\gamma_2}} > 35\,\,\GeV, \nonumber \\
\quad\quad & \qquad |y_{\gamma}| < 2.5, \quad\quad
\Delta R_{\gamma_1\gamma_2} > 0.4, \;
\end{align}
which are designed to reduce sensitivity on infrared
physics~\cite{Salam:2021tbm}.
The cuts in Eq.~\eqref{eq:fiducialsetup} are different from the ones
used in Refs.~\cite{Dixon:2013haa,deFlorian:2013psa}. Because of this,
it is not immediate to compare our results with the ones presented
there. Nevertheless, we have validated our LO and NLO calculations
against Refs.~\cite{Dixon:2013haa,deFlorian:2013psa}. In particular,
we have reproduced the results for the mass shift shown in
Refs.~\cite{Dixon:2013haa,deFlorian:2013psa}. We have also 
used \texttt{MCFM}~\cite{Campbell:1999ah,Boughezal:2016wmq,Campbell:2019dru}, 
together with in-house
codes, to validate the signal and the reliability of the $\NNLOsvp$ approximation.

We start the discussion of our results by assessing the validity of
the soft-virtual approximation for the interference at NLO.  We do
this by first comparing the exact prediction including only the $gg$
initiated process, which we dub $\NLO(\mathrm{gg})$, to the
corresponding soft-virtual approximation, $\NLOsv(\mathrm{gg})$.
\begin{figure*}[t!]\begin{center}\begin{tabular}{ccc} 
  \hspace{-0.9cm}
  \includegraphics[scale=0.35,trim=50 50 90 50]{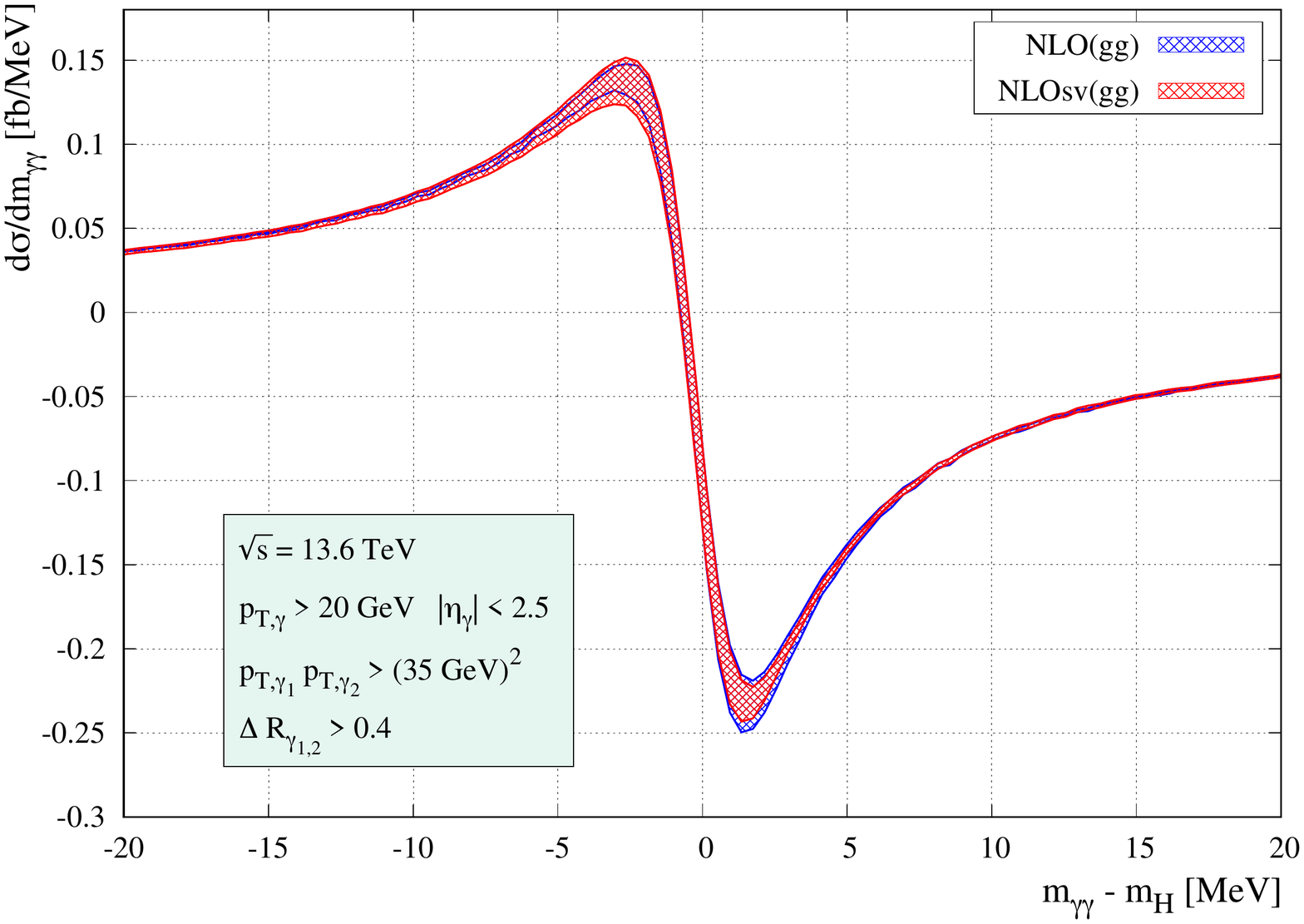} & & 
  \includegraphics[scale=0.35,trim=50 50 90 50]{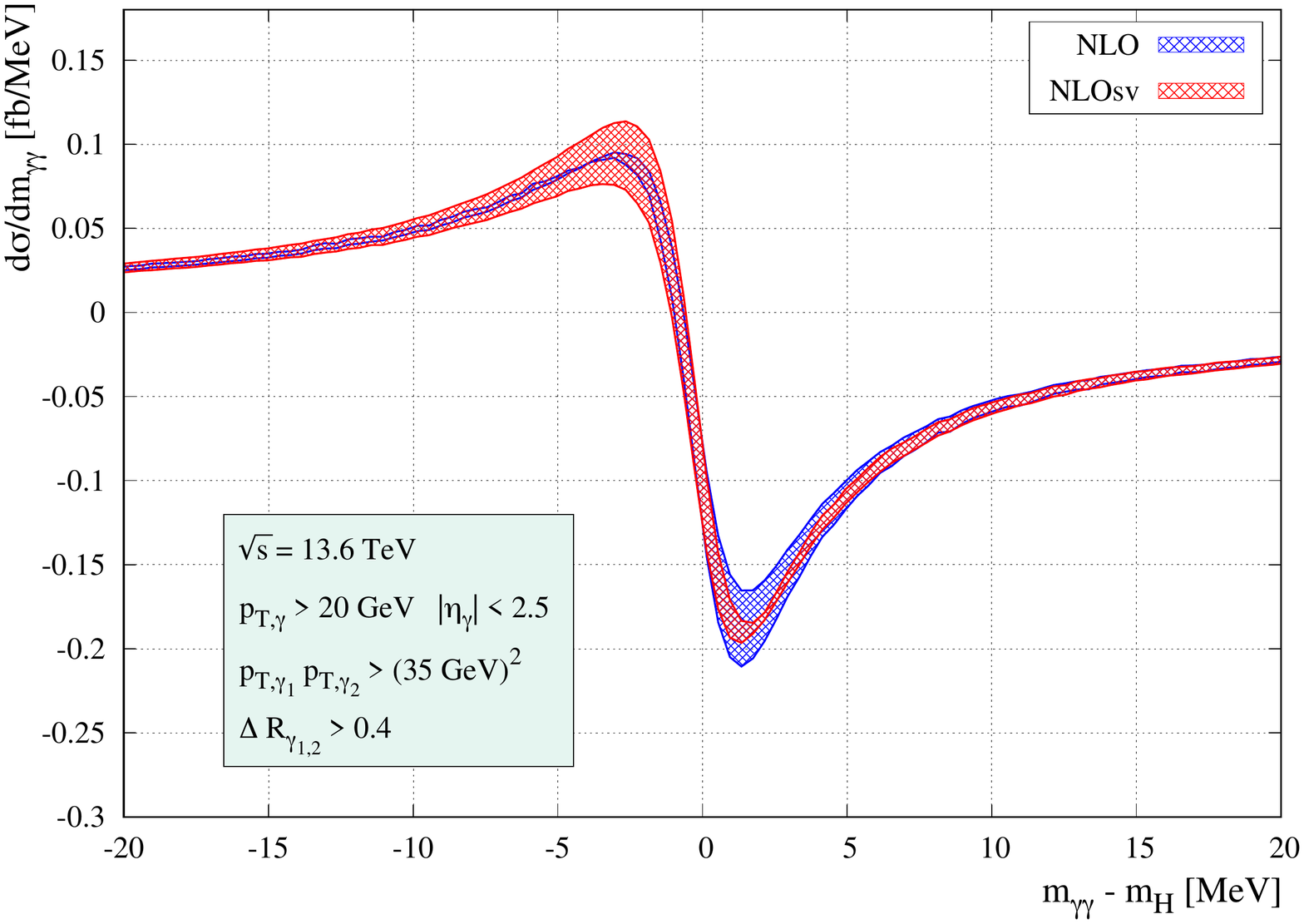}\vspace{-0.4cm}
  \end{tabular}\end{center}
  \caption{Left pane: comparison of the exact NLO calculation and the soft-virtual approximation in the $gg$ channel.
  Right pane: complete NLO prediction, inclusive of all channels, compared to
  the corresponding soft-virtual approximation}\label{fig:validationsv} 
\end{figure*}
This comparison is shown in the left pane of
Fig.~\ref{fig:validationsv}.  Second, we compare the complete NLO
calculation, \ie including the $qg$ and $\bar{q}g$ initiated channels~\footnote{We
  note that the impact of the $q\bar q$ channel is negligible.} to the NLO soft-virtual
approximation ($\NLOsv$), which retains all LO contributions, but 
includes NLO corrections in the soft-virtual
approximation for the $gg$ channel only.
This
comparison is presented in the right pane of
Fig.~\ref{fig:validationsv}.  The red and blue bands shown in both
figures represent the envelope from a simultaneous rescaling of
$\mu_{F,R}$ by a factor of two up and down with respect to the central
value.
We note that our result suggests that the uncertainty due to scale variations does not provide a
reliable estimate of the actual error of the NLO prediction.
In the left pane of
Fig.~\ref{fig:validationsv} one can see that the soft-virtual
approximation does a remarkably good job in describing the dominant
$gg$ contribution.  The shape of the interference is very well
captured and the scale-variation bands almost perfectly match
throughout the relevant interval of invariant mass.  In the right side
of Fig.~\ref{fig:validationsv} we notice a slight degrading of the
approximation, mostly for what concerns the scale variation
bands. This is because the soft-virtual approximation does not capture
the full scale-compensation between different channels that happens
at NLO. 
However, the shape, which is the relevant factor for the
extraction of the mass-shift, is still described at a 
satisfactory degree.

This comparison gives us some confidence that the soft-virtual
approximation can adequately describe higher-order QCD corrections
to the signal-background interference. 
Furthermore, we note that the
mass-shift extracted at NLO in the soft-virtual approximation differs
from the exact value by $5\%$, whilst the genuine NLO correction
is of $\calO(30/40\%)$ and the theory uncertainty is
$\calO(10\%)$. The ability of the soft-virtual approximation in
reproducing the interference effects can be better understood by
keeping in mind that the predominant contributions come from the low
$p_{T,\gamma\gamma}$ regions, as we already mentioned in
Sec.~\ref{intro}. We point out that the fiducial setup adopted
in Eq.~(\ref{eq:fiducialsetup}), in particular with the choice of
product cuts, plays a relevant role in the reliability of the
soft-virtual approximation.  We have indeed explicitly checked that
imposing asymmetric cuts on the final-state photons, as it is routinely
done in Higgs analysis, breaks the quality of the approximation.

\begin{figure}[t!]
  \begin{center}
  \hspace{-0.9cm}
  \includegraphics[scale=0.34,trim=68 50 88 50]{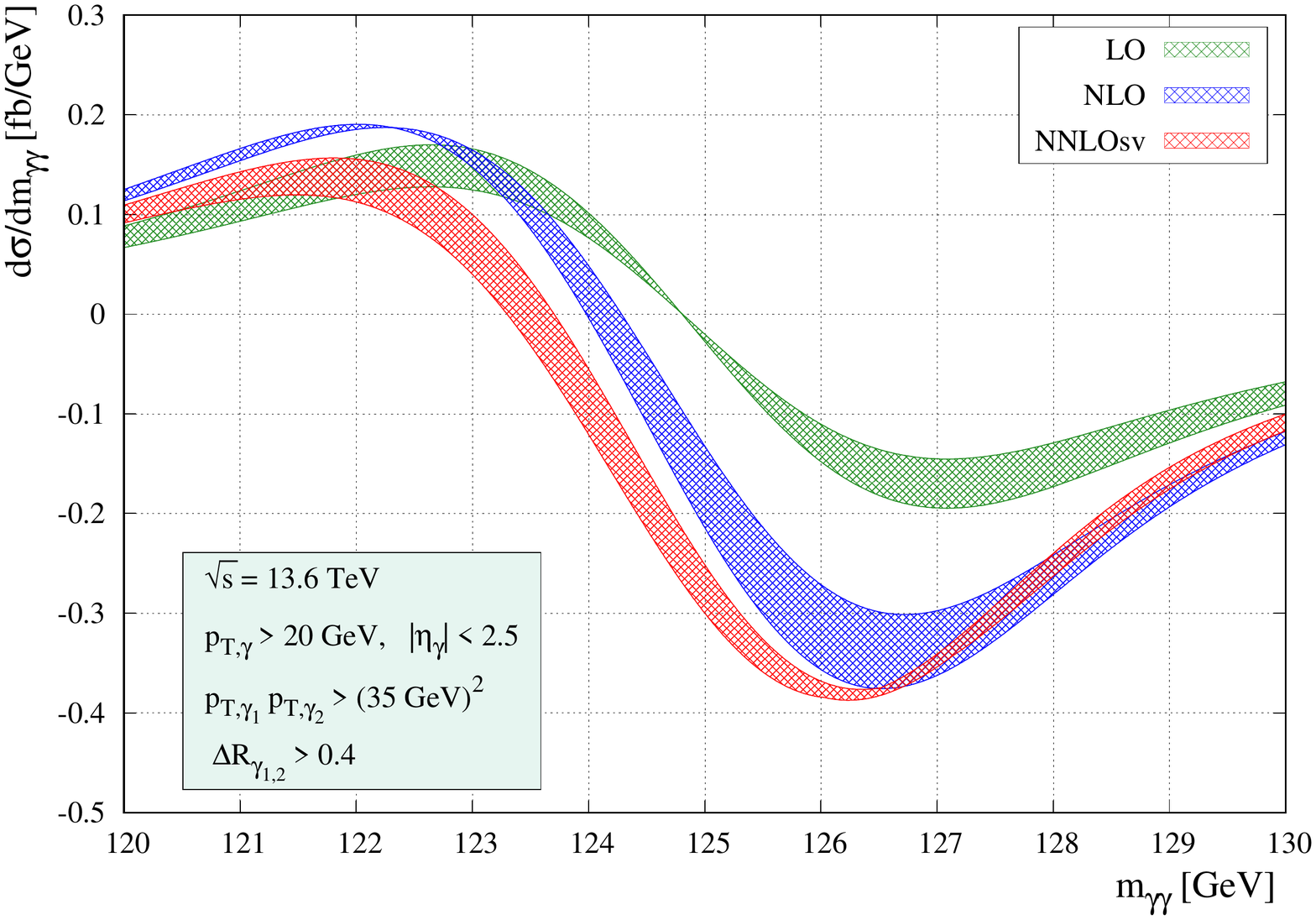}
  \caption{Signal-background interference contribution to the diphoton invariant mass distribution after 
  Gaussian smearing. Bands represent the envelope given by 
  the scale variation.}
  \label{fig:interfnnlo}
  \end{center}
\end{figure}
As already discussed in Sec.~\ref{se:mass_shift},
in order to get a more realistic picture of the interference effect at
the detector level, we convolute the invariant mass distribution with
a Gaussian function with a standard deviation of $\sigma =
1.7~\GeV$~\cite{Martin:2012xc,ATLAS:2016kvj}

The invariant mass distribution
arising from signal-background interference after the smearing is
shown in Fig.~\ref{fig:interfnnlo}. The LO features the well known
antisymmetric behaviour around the peak coming from the real part of
the interference, whereas the NLO curve is shifted to the left and to
the bottom, due to the impact of the imaginary part.
In the same figure, we can appreciate for the first time the effect of
the NNLO QCD corrections. The curve is further shifted down, thus
depleting even more the number of events around the Higgs peak and
softening the impact on the mass-shift.
\begin{figure*}[t!]\begin{center}\begin{tabular}{ccc} 
  \hspace{-0.3cm}
  \includegraphics[scale=0.35,trim=70 50 90 50]{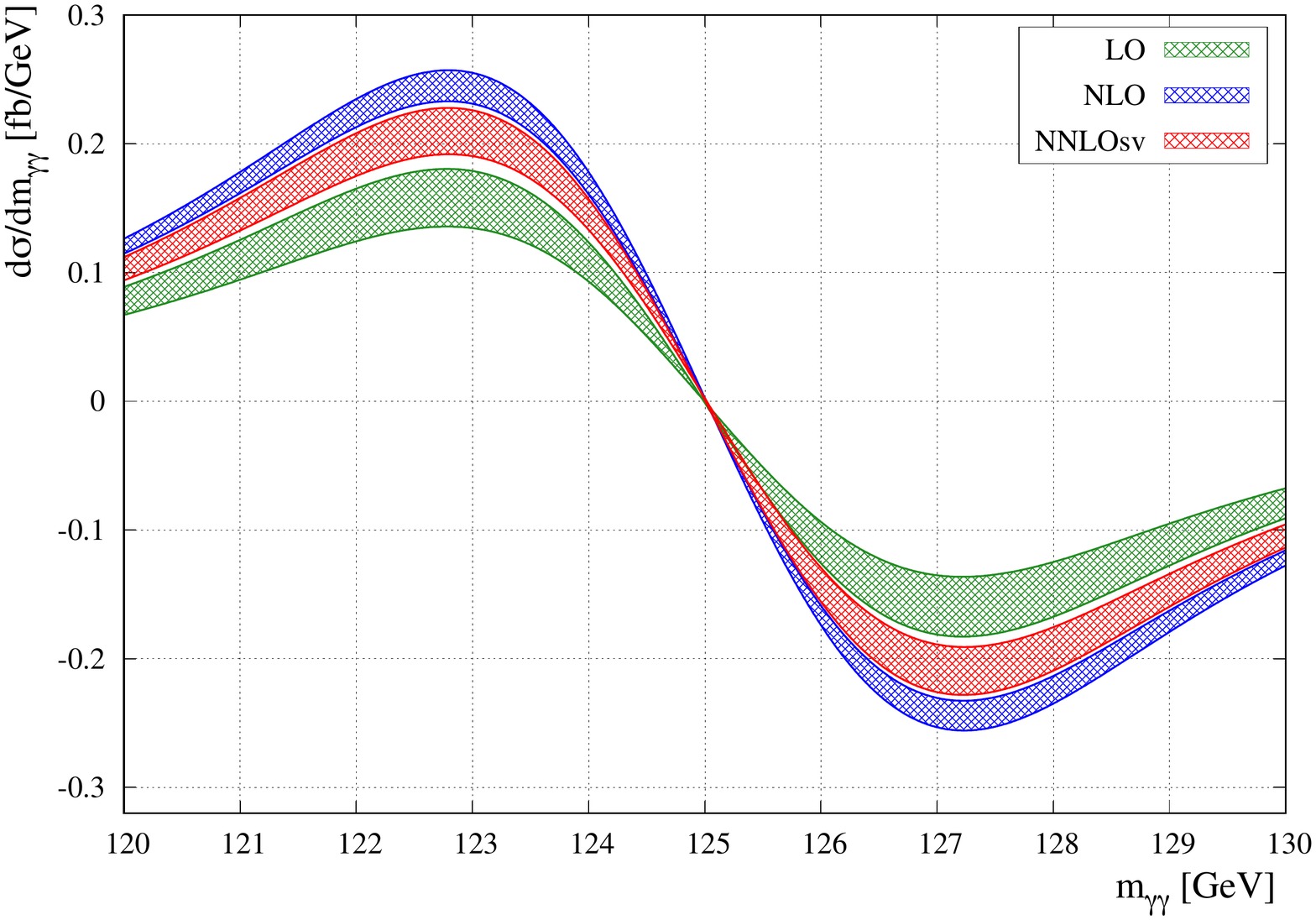} & & \hspace{-0.2cm}
  \includegraphics[scale=0.35,trim=50 50 90 50]{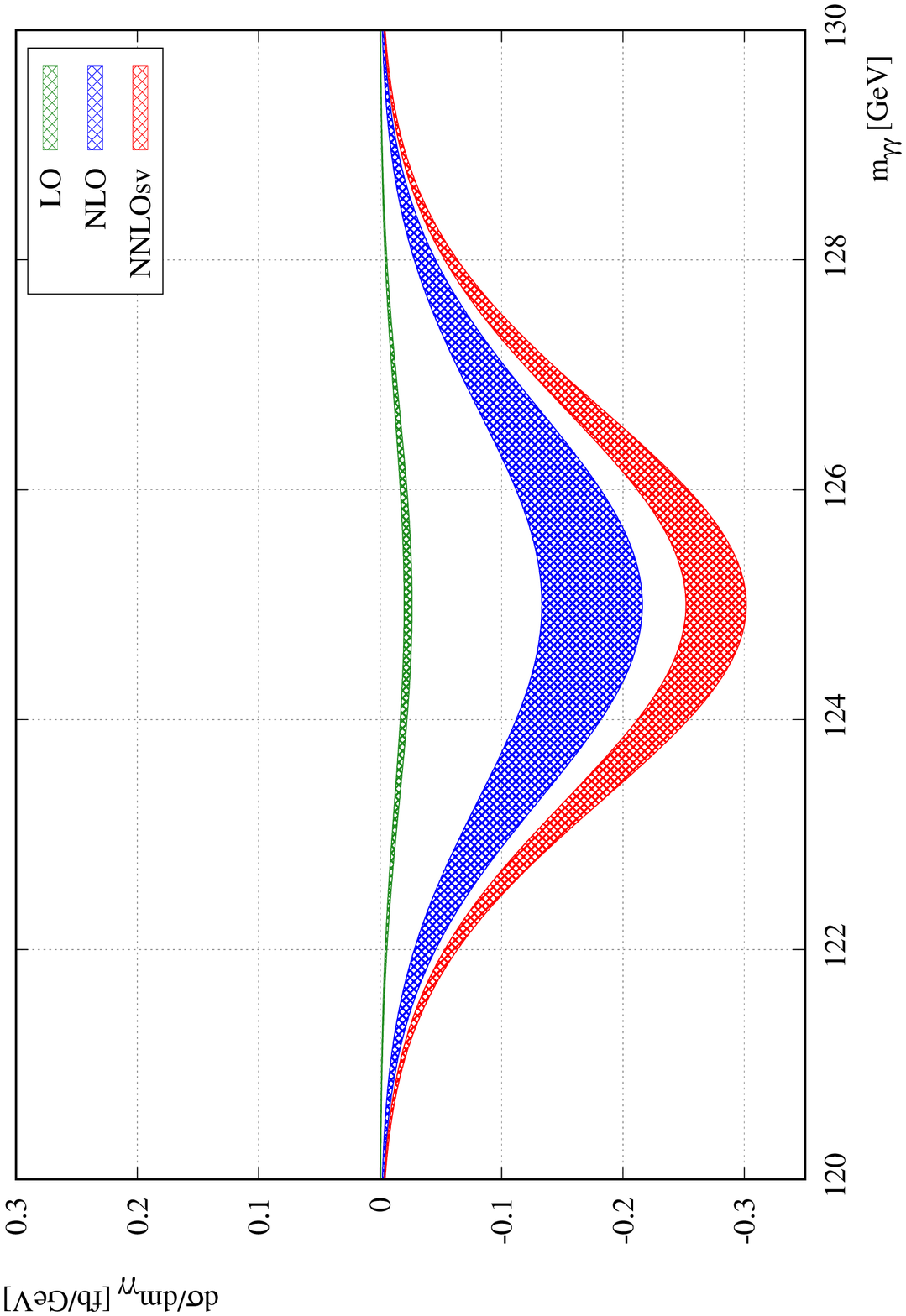} \vspace{-0.5cm}
  \end{tabular}\end{center}
  \caption{Real part (left pane) and imaginary part (right pane) of the interference at LO, NLO and NNLOsv
  after Gaussian smearing.}\label{fig:reimsmeared} 
\end{figure*}
Moreover, from Fig.~\ref{fig:interfnnlo} one can see that the theory uncertainty, 
which arises almost completely from renormalisation scale variations, 
gets generally reduced at NNLO, except in the region where the NLO band 
shrinks to zero.
This behaviour can be understood keeping in mind that the interference
is the sum of the real and imaginary parts which have different shapes, as discussed in Sec.~\ref{se:higgsint}.
As shown in Fig.~\ref{fig:reimsmeared}, scale-variation
bands are separately well behaved for the real and imaginary parts,
but on the left hand side of the Higgs boson peak a cancellation occurs,
whereas on the right the two effects sum up (both negative in sign).

Before moving to our results for the mass shift, we briefly 
discuss the impact of the imaginary part of the interference to the
total cross section in our fiducial region. As we have explained in
Sec.~\ref{se:higgsint}, at LO this interference is very small because
it only comes from contributions mediated by virtual bottom quarks,
either in the signal or in the background. These are either
mass (background) or Yukawa (signal) suppressed. The smallness of the LO
imaginary part is indeed seen in Fig.~\ref{fig:reimsmeared}. In our setup,
we find
\begin{equation}
  \sigma_S^{\LO} = 24.21^{+15\%}_{-14\%}~\fb,~~~~~~ \sigma_I^\LO =
  -0.11^{+20\%}_{-17\%}~\fb.
\end{equation}
Here and in the following the quoted uncertainties are obtained by
coherently varying the renormalisation and factorisation scales by
a factor of two around the central value $\mu = m_{\gamma\gamma}/2$. 
At LO, we find that more than 80\% of the destructive interference
quoted above comes from the imaginary part of the signal interfering
with the real part of the background. This gives us confidence that
neglecting mass effects in the background prediction does not significantly impact
our result. Furthermore, as far as the signal goes, we note that the bulk
(about 95\%)
of the imaginary part is generated by bottom-mass effects in the production
amplitude. This is easy to understand just by looking at the relative
importance of the top, bottom and $W$ contributions to the production and
decay amplitudes. 

At higher orders however, a larger interference is generated by the
imaginary part of the background, which no longer requires the presence
of bottom quarks (see the discussion in Sec.~\ref{se:highordcorr}).
Because of this, beyond LO we
only compute radiative corrections in the infinite-top approximation
and drop any mass dependence in the background amplitudes. At NLO, we
obtain
\begin{equation}
\sigma_S^\NLO = 58.12^{+20\%}_{-14\%}~\fb,~~~~~~ \sigma_I^\NLO =
-0.72^{+27\%}_{-21\%}~\fb.
\end{equation}
These results are consistent
with the analysis in Ref.~\cite{Campbell:2017rke}. Our best prediction
beyond NLO is obtained within the soft-virtual approximation described in
Sec.~\ref{se:highordcorr}. We find
\begin{equation}
\label{eq:xsection}
\sigma^{\NNLOsv'}_{S} = 72.21^{+8\%}_{-8\%}\; \mathrm{fb}, ~~~~
\sigma^{\NNLOsv}_{I} = -1.21^{+7\%}_{-10\%}\; \mathrm{fb},
\end{equation}
hence the destructive
interference reduces the total rate by $1.7\%$.\footnote{We point out that 
the theory uncertainties for the signal cross section in Eq.~\eqref{eq:xsection}
have been computed employing the exact NNLO QCD scale variations.}  Given the
theoretical~\cite{LHCHiggsCrossSectionWorkingGroup:2016ypw} (see also
Refs.~\cite{Giulia,Robert}) and
experimental~\cite{ATLAS:2022fnp,CMS:2022wpo} uncertainty on the Higgs
total cross section, this effect is actually not negligible and it can
be used to further constrain the Higgs width~\cite{Campbell:2017rke}.
We do not pursue this line of investigation here, but we estimate that,
with current uncertainties, one could already constrain the Higgs width
to about 20-30 times the Standard Model.

\begin{table}[t]
  \begin{center}
  \begin{tabular}{|c|c|c|c|}
  \hline
  $\Delta m_{\gam\gam}\left[\text{MeV}\right]$ & 7 TeV                     & 8 TeV                     & 13.6 TeV                 \\ \hline\hline
  $\LO$                                        & $-77.2_{-1.0\%}^{+0.8\%}$ & $-79.5_{-0.8\%}^{+0.6\%}$ & $-83.1^{+0\%}_{-0.3\%}$         \\ \hline
  $\NLO$                                       & $-56.2_{-15\%}^{+13\%}$   & $-56.8_{-14\%}^{+13\%}$   & $-55.2_{-12\%}^{+12\%}$  \\ \hline
  $\NNLOsv$                                    & $-46.3_{-17\%}^{+15\%}$   & $-47.0_{-16\%}^{+14\%}$   & $-46.0_{-12\%}^{+11\%}$  \\ \hline
  $\NNLOsvp$                                   & $-39.5_{-24\%}^{+20\%}$   & $-39.7_{-22\%}^{+19\%}$   & $-39.4_{-17\%}^{+16\%}$ \\ \hline
  \end{tabular}
  \caption{Mass-shift at different proton-proton collider energies with Gaussian fit method.}
  \label{tab:gaussianshift}
  \end{center}
  \begin{center}
  \begin{tabular}{|c|c|c|c|}
  \hline
  $\Delta m_{\gam\gam}\left[\text{MeV}\right]$ & 7 TeV                      & 8 TeV                      & 13.6 TeV                   \\ \hline\hline
  $\LO$                                        & $-113.4_{-1.0\%}^{+0.8\%}$ & $-116.7_{-0.8\%}^{+0.6\%}$ & $-122.1_{-0.3\%}^{+0.1\%}$ \\ \hline
  $\NLO$                                       & $-82.6_{-15\%}^{+13\%}$    & $-82.8_{-14\%}^{+12\%}$    & $-81.2 _{-12\%}^{+12\%}$   \\ \hline
  $\NNLOsv$                                    & $-68.1_{-17\%}^{+15\%}$    & $-68.4_{-15\%}^{+13\%}$    & $-67.7 _{-12\%}^{+11\%}$   \\ \hline
  $\NNLOsvp$                                   & $-58.1_{-23\%}^{+20\%}$    & $-59.2_{-21\%}^{+18\%}$    & $-58.0 _{-17\%}^{+16\%}$  \\ \hline
  \end{tabular}
  \caption{Mass-shift at different proton-proton collider energies with first moment method.}
  \label{tab:momentshift}
  \end{center}
\end{table}
We can finally present the main result of our study, \ie the
prediction for the mass-shift at NNLO.  As discussed in
Sec.~\ref{se:higgsint}, we adopt two different methods to estimate the
mass-shift induced by the interference term.  In
Table~\ref{tab:gaussianshift} we show the results obtained by
performing a chi-squared fit of the smeared signal-plus-interference
distribution with a Gaussian function of standard deviation
$\sigma=1.7\,\GeV$. The mass-shift is obtained as the difference
between the obtained mean value and the input Higgs boson mass $m_H =
125~\GeV$.  In Table~\ref{tab:momentshift} instead we present results
derived by computing the first moment of the signal-plus-interference
distribution after smearing.  In both tables we show predictions for
different collider energies, but in the same fiducial region specified
in Eq.~(\ref{eq:fiducialsetup}). As mentioned earlier, the
denoted ranges represent the theoretical uncertainty related to a
change of the central scale $\mu=m_{\gamma\gamma}/2$ by a factor of two and a half.
The entry $\NNLOsv$ indicates the result obtained by considering both
signal and interference terms in the soft-virtual approximation. The
$\NNLOsvp$ entry instead, refers to the ``improved'' soft-virtual
approximation discussed at the end of Sec.~\ref{se:highordcorr}.
Specifically, in this case we still use the $\NNLOsv$ approximation
for the interference, but compute the signal in the $\NNLOsvp$ framework.
As we explained in Sec.~\ref{se:highordcorr}, we expect this setup to
be the most realiable one. Still, we find it useful to present numbers
in both frameworks as a conservative way of estimating the uncerainties
related to these approximations. In this respect, we note that the results
obtained in the $\NNLOsv$ and $\NNLOsvp$ approximations are compatible
 within their uncertainties. 

We immediately notice that in both extraction methods,
the estimated mass-shift is rather insensitive to the collider energy.
\begin{table}[t]
  \begin{center}
  \begin{tabular}{|c|c|c|c|}
  \hline
  $\Delta m_{\gam\gam}\left[\text{MeV}\right]$ & First moment  & Gaussian Fit \\ \hline\hline
  $K_{\NLO}$                                   & 0.665         & 0.664        \\ \hline
  $K_{\NNLOsv}$                                & 0.554         & 0.554        \\ \hline
  $K_{\NNLOsvp}$                               & 0.475         & 0.474        \\ \hline
  \end{tabular}
  \caption{Comparison of $K$-factors, measured \wrt the LO value,
    for the mass-shift at $\sqrt{s}=13.6~\TeV$ calculated via a gaussian fit method and via a first-moment method.}
  \label{tab:kfact}
  \end{center}
\end{table}
Although a Gaussian fitting procedure and the evaluation of the first moment
return different values for the mass-shift, smaller for the former and larger
for the latter, we notice that the $K$-factors relative to the corresponding
LO prediction are almost identical. We stress that this is a welcome feature
since our analysis does not include reliable
detector simulation. The results in Table~\ref{tab:kfact} show that radiative
corrections seem rather insensitive to the precise definition of the observable
used to extract the mass shift. 

We also observe how the mass-shift gets systematically reduced in absolute
value when including higher-order corrections.
This is mainly because the $K$-factor for the signal turns out to be
larger than the one for the interference, thus leading to a reduction of
the mass-shift. This trend, already observed at NLO~\cite{Dixon:2013haa},
seems to persist at NNLO (at least in the $\NNLOsv$ approximation).

\begin{figure}[t!]
  \begin{center}
  \hspace{-0.9cm}
  \includegraphics[scale=0.34,trim=68 50 88 50]{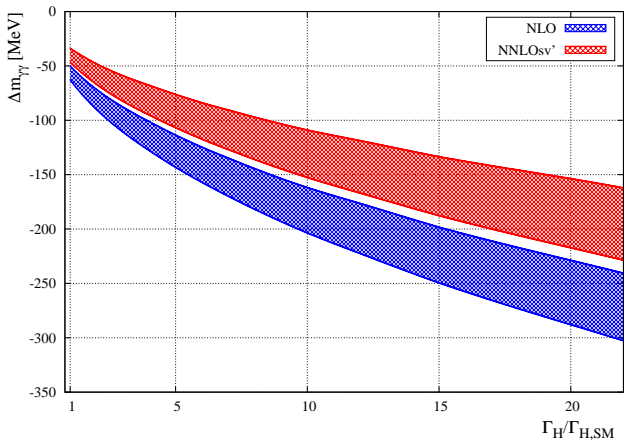}
  \caption{Mass-shift as a function of the Higgs boson width}
  \label{fig:shiftwidth}
  \end{center}
\end{figure}

Finally, we study how the bounds on $\Gamma_H$ are affected by
higher-order QCD corrections, following the discussion at the end of
Sec.~\ref{se:higgsint}.  The results at NLO and $\NNLOsvp$ (i.e. our
best prediction) are
shown in Fig.~\ref{fig:shiftwidth}.  Our NLO curve is compatible with the one
obtained in~\cite{Dixon:2013haa}.\footnote{We remind the reader that
  our results are derived for a different collider energy and fiducial
  cuts.} We note that the two curves
do not overlap. This is a direct consequence of the analogous feature
for the interference shown in Fig.~\ref{fig:interfnnlo}. As explained
in the comment of that figure, this is at least partially due to 
competing effects coming from real and imaginary part of the interference,
see Fig.~\ref{fig:reimsmeared}. Furthermore, similar to the Higgs-signal
case, the interference receives rather large corrections as well. 
Finally, we note that the NNLO soft-virtual curve lies above the
NLO one, thus loosening potential bounds on $\Gamma_H$. For instance,
if at the LHC the error on the mass-shift reaches roughly $150~\MeV$,
one could exclude values of $\Gamma_H$ larger than 10--20 times the SM
prediction.  This has to be compared with the corresponding bound at
NLO of $\Gamma_H<(6\text{--}9)\Gamma_{H,\SM}$. 

\section{Conclusions} \label{se:conclusions}
We presented a calculation of the NNLO soft-virtual corrections to
signal-background interference for gluon-fusion Higgs production in the
diphoton channel at the LHC. More specifically, we focused
on the shift in the diphoton invariant mass distribution which arises
from the inclusion of the interference between the Higgs-mediated
process and its continuum background, as first observed in
Ref.~\cite{Martin:2012xc}.  Such a study was only known up to NLO QCD
so far~\cite{Dixon:2013haa}, due to unavailability of the relevant
background amplitudes.

Recent developments in multi-loop scattering amplitudes techniques
enabled the computation of such
amplitudes~\cite{Bargiela:2021wuy,Agarwal:2021vdh,Badger:2021imn}. We
were therefore able to extend this study one order higher in
perturbation theory.  We employed the soft-virtual approximation,
motivated by previous studies which showed how the contribution to the
mass-shift is enhanced in the low $p_{T,\gam\gam}$ region, and
therefore dominated by virtual corrections and soft emissions. We have
validated this approximation at NLO and shown that indeed it seems to
capture the bulk of the corrections, to within few
percent.
We stress that this statement holds in our fiducial
volume, which is designed to avoid spurious sensitivity to infrared
physics~\cite{Salam:2021tbm}.
Using more standard asymmetric cuts at fixed order, instead, would lead to a deterioration of the quality of the soft-virtual approximation.

We calculated the mass-shift by means of two different methods and
found that, despite the absolute value being dependent on the observable
chosen for its extraction, the $K$-factors for the NNLO soft-virtual
corrections are very similar in both cases. Specifically, we found
that the mass-shift further decreases at NNLO soft-virtual, confirming
the trend already observed at NLO QCD. Within our approximation, we
find that NNLO corrections are large, and amount to a decrease of
almost 30\% on top of the NLO prediction.
Furthermore, these corrections are not captured by the standard scale variation band of the NLO prediction.

We used our results to obtain an updated prediction for the bounds
that can be put on $\Gamma_H$ as a result of the mass-shift
determination. We observed that the decrease in the mass-shift
prediction results in a loosening of such bounds. If we estimate the
experimental error on the mass shift to drop to $\calO(150~\MeV)$, we
find that the Higgs boson width can be constrained to $\Gamma_H\lesssim
(10-20)\Gamma_{H,SM}$. Interestingly, we find that the imaginary part
of the interference instead grows with the perturbative order. Hence,
it leads to a shift of the total cross section that could be used to
extract information on the Higgs total width as well~\cite{Campbell:2017rke}.
Although a detailed extraction based on this is beyond the scope of this
paper, we estimate that with the current experimantal and theoretical
uncertainties one could bound the Higgs width to within 20-30 times its
Standard Model value. 

There are several avenues for improving our predictions. First, one
could perform an exact NNLO study. As we already mentioned, the main
complication of such a calculation is a purely technical one, namely
the numerical control of two-loop 5-point amplitudes (as well as
6-point one-loop ones) near singular contributions. Although we believe
that our calculation captures the bulk of NNLO corrections, such
an exact result would allow for a better modeling of the $p_T$-dependence
of the interference, which would provide useful information for an actual extraction of the mass-shift. 
Indeed, the $p_T$ dependence of the result
can be used to define signal and control regions to extract the mass shift
within the diphoton channel alone thus minimising systematic uncertainties. 

From a theoretical point of view, perhaps an even more interesting line
of investigation would be to understand how to improve the soft-virtual
approximation for the background amplitude. This would require improving
our understanding of next-to-leading power soft corrections to cope
with the case at hand. We leave these avenues for future work. 

\section*{Acknowledgments}
We would like to thank M. van Beekveld for
discussions on the next-to-leading power soft approximation and
T. Neumann for correspondence on MCFM.
FD wishes to thank K. Melnikov and C. Signorile-Signorile for
useful discussions.
This research was supported by
ERC
Starting Grants 804394 \textsc{HipQCD} (PB, FB, FC, FD) and
949279 HighPHun (LT), by the UK Science and
Technology Facilities Council (STFC) under grant ST/T000864/1 (FC)
and  by the Excellence Cluster ORIGINS funded by
the DFG under Germany’s Excellence Strategy - EXC-2094 - 39078331 (FB, LT).
AvM was supported in part by the
National Science Foundation (NSF) through Grant 2013859.
FC, FB and LT also acknowledge support by the Munich Institute for
Astro-Particle and BioPhysics (MIAPbP) which is funded by the Deutsche
Forschungsgemeinschaft (DFG, German Research Foundation) under
Germany's Excellence Strategy – EXC-2094 – 390783311. FD would like to
thank the Karlsruhe Institute of Technology for hospitality during the
last stages of this work.

\appendix

\section{Details on the signal amplitude} \label{app:amp_def}
In this appendix we provide some details regarding our treatment of the
Higgs signal amplitudes.
For the decay component of the amplitude, 
we retain full dependence on the heavy quark masses at LO and neglect higher order radiative corrections. More specifically, we consider both 
quark and W-boson loops contributions,
\begin{align}
\label{eq:Haadecay}
&\mathcal{A}_{H \gamma\gamma} = \\ \nonumber  
&-\ff{\alpha_{em}}{4\pi v}
m_{\gam\gam}^2\Big[I_W\left(\ff{m_W^2}{m_{\gam\gam}^2}\right)+\sum_{q=b,t}
N_c Q_q^2 I_q\left(\ff{m_q^2}{m_{\gam\gam}^2}\right)
\Big]
\end{align}
where $N_c$ is the number of colors in the fundamental representation of SU(3),
$Q_q^2$ is the quark electric charge in units of the proton charge
and 
\begin{align}
  I_W(x) &=-2[6x+1+6x(2x-1)F(x)] \nonumber \\
  I_q(x) &= 4x[2+(4x-1)F(x)] \nonumber \\
  F(x) &=
  \theta(1-4x)\ff{1}{2}\Big[\log\left(\ff{1+\sqrt{1-4x}}{1-\sqrt{1-4x}}
  \right)-i\pi\Big]^2 \nonumber \\ 
  &-\theta(4x-1)2\left(\sin^{-1}\left(\ff{1}{2\sqrt{x}}\right)\right)^2\,.
\end{align}

As mentioned in Sec.~\ref{se:highordcorr}, Higgs production
is treated exactly at LO while the heavy-top effective field 
theory is employed at higher orders. In the case of exact top and bottom
mass dependence, the expression for the production amplitude is
\begin{equation}
\label{eq:Hggprod}
\mathcal{A}_{ggH} =-\ff{\alpha_{s}}{4\pi v}
\frac{m_{\gam\gam}^2}{2} \,\sum_{q=b,t} Q_q^2 I_q\left(\ff{m_q^2}{m_{\gam\gam}^2}\right)\, .
\end{equation}
In the heavy-top effective theory, 
the Higgs production amplitude is described by the Lagrangian in Eq.~\eqref{eq:eff_lag}.
The effective coupling $\lambda$ reads
\begin{align}\label{eq:lambdahighord}
  &\lambda = -\ff{\as}{3\pi v}C(\as)Z_{\lambda}(\as)\,,
\end{align}
with $Z_{\lambda}(\as)$ and $C(\as)$ in the $\MSbar$ scheme given by 
\begin{align}
Z_{\lambda}(\as) &= 1-\ff{\beta_0}{\ep}\left(\ff{\as(\mu)}{2\pi}\right)
+\Big[\ff{\beta_0^2}{\ep^2}-\ff{\beta_1}{\ep}\Big]\left(\ff{\as(\mu)}{2\pi}\right)^2 \nonumber \\ 
&+ \calO(\as^3), \\
C(\as) &= 1+\Big[\ff{5}{2}C_A-\ff{3}{2}C_F\Big]\asontwopi+\Big[\ff{1063}{144}C_A^2 \nonumber \\
&-\ff{25}{3}C_A C_F + \ff{27}{8}C_F^2 - \ff{47}{72}C_A n_f - \ff{5}{8}C_F n_f \nonumber \\
&-\ff{5}{48}C_A -\ff{C_F}{6}+\log\ff{\mu^2}{m_t^2}\Big(\ff{7}{4}C_A^2 -\ff{11}{4}C_A C_F \nonumber \\
& + C_F n_f\Big)\Big]\asontwopi^2 + \calO(\as^3),
\end{align}  
see e.g.\ Ref.~\cite{Grigo:2014jma}.
In the equation above we use
\begin{align}
  &\beta_0 = \ff{11}{6}C_A-\ff{2}{3}n_fT_R \nonumber \\
  &\beta_1 = \ff{17}{6}C_A^2-\ff{5}{3}C_A T_R n_f -C_F T_R n_f
\end{align}
and $\as(\mu)$ is the renormalized coupling.
The 1-loop and 2-loop gluon form factors which are relevant for the 
calculation of NLO and NNLO corrections to the signal amplitude have been 
taken from Ref.~\cite{Gehrmann:2005pd}.

\section{Soft-virtual cross-section at NLO and NNLO} \label{app:NNLOsv}
In Section \ref{se:highordcorr} we saw that the soft-virtual approximation
consists in retaining only soft and virtual contributions to the cross-section
and neglect hard and collinear contributions. In this appendix, we will define 
the various virtual contributions and present formulas for the 
soft-virtual cross-section for color singlet production at NLO and NNLO.
As mentioned in Section \ref{se:highordcorr}, the soft-virtual cross-section
is completely universal for a given partonic channel. We will present results for the $gg$-channel,
since it is the one of interest for this work. The only 
process-dependent component is encoded in the virtual contributions.
The NLO virtual contribution to the cross-section can be written as 
\begin{align}\label{eq:NLOvirt}
\hat{\sigma}^{\mathrm{1-loop}} = \ff{\as(\mu)}{2\pi} I_{12}(\ep) \hat{\sigma}_{\mathrm{Born}} + 
\hat{\sigma}^{\mathrm{1-loop}}_{\mathrm{fin}}
\end{align}  
where $\hat{\sigma}^{\mathrm{1-loop}}_{\mathrm{fin}}$ is the  contribution to the cross section due to the interference of the finite remainder 
of the one-loop amplitude with the Born amplitude, and $I_{12}(\ep)$ is the Catani operator
\begin{equation}
I_{12}(\ep) = -2\cos(\pi \ep)\ff{e^{\ep \gamma_E}}{\Gamma(1-\ep)}\left(\ff{\mu^2}{s_{12}}\right)^{\ep}
\left[\ff{C_g}{\ep^2}+\ff{\gamma_g}{\ep}\right]
\end{equation}
with $C_g = C_A$ and $\gamma_g = \beta_0$.
The NLO soft-virtual cross section then reads
\begin{align}
  \label{eq:NLOsv}
  &G^{(1)}(z) = 
  \delta(1-z)\left(\ff{2}{3}\pi^2+\ff{\hat{\sigma}^{\mathrm{1-loop}}_{\mathrm{fin}}}{\hat{\sigma}_{\mathrm{Born}}}
    -2\beta_0 \logs\right) \nonumber \\
  &\qquad  +8 C_A \mathcal{D}_1(z) -4 C_A \mathcal{D}_0(z)\logs  +\mathcal{O}(\ep).
\end{align} 
At NNLO, the virtual contribution can be written as 
\begin{align}\label{eq:NNLOvirt}
&\hat{\sigma}^{\mathrm{2-loop}} = \left(\ff{\as(\mu)}{2\pi}\right)^2 \hat{\sigma}_{\rm{Born}} \Big[\ff{I_{12}^2(\ep)}{2} - \ff{\beta_0}{\ep} 
I_{12}(\ep) 
\nonumber \\ &
+ e^{-\ep \gamma_E} \ff{\Gamma(1-2\ep)}{\Gamma(1-\ep)}\left(\ff{\beta_0}{\ep}+K\right)I_{12}(2\ep) + 
\ff{e^{\ep \gamma_E}}{\Gamma(1-\ep)}\ff{H_g}{\ep}\Big]
\nonumber \\ &
+ \left(\ff{\as(\mu)}{2\pi}\right) \hat{\sigma}^{\mathrm{1-loop}}_{\mathrm{fin}} I_{12}(\ep) 
+\hat{\sigma}^{\mathrm{1-loop^2}}_{\mathrm{fin}} + \hat{\sigma}^{\mathrm{2-loop}}_{\mathrm{fin}}.
\end{align}
where
\begin{align}
 K &= \left(\ff{67}{18}-\ff{\pi^2}{6}\right)C_A - \ff{10}{9}T_R n_f  \\
 H_g &= C_A^2 \left(\ff{5}{12} + \ff{11}{144}\pi^2 + \ff{\zeta_3}{2} \right)
+C_A\,n_f\left(-\ff{29}{27}-\ff{\pi^2}{72}\right)
\nonumber \\ &
+ \ff{C_F\,n_f}{2}+\ff{5}{27}n_f^2. 
\end{align}
Furthermore, $\hat{\sigma}^{\mathrm{1-loop}}_{\mathrm{fin}}$ is defined in Eq.~\eqref{eq:NLOvirt},
$\hat{\sigma}^{\mathrm{1-loop^2}}_{\mathrm{fin}}$ and $\hat{\sigma}^{\mathrm{2-loop}}_{\mathrm{fin}}$
are implicitly defined in Eq.~\eqref{eq:NNLOvirt} and represent the contributions due to the 1-loop finite remainder squared
and the interference of the 2-loop finite remainder with the Born amplitude, respectively.

The NNLO soft-virtual cross section then reads
\begin{align}
  \label{eq:NNLOsv}
&G^{(2)}(z) =\ff{\alpha_s(\mu)}{2\pi}\Bigg\{\ff{\hat{\sigma}^{\mathrm{2-loop}}_{\mathrm{fin}}}{\hat{\sigma}_{\mathrm{Born}}}
+ \ff{\hat{\sigma}^{\mathrm{1-loop^2}}_{\mathrm{fin}}}{\hat{\sigma}_{\mathrm{Born}}} +
\ff{\hat{\sigma}^{\mathrm{1-loop}}_{\mathrm{fin}}}{\hat{\sigma}_{\mathrm{Born}}} \times \nonumber \\ 
&\Bigg[\delta(1-z)
\left(\ff{2}{3}\pi^2 C_A - 2\beta_0 \logs\right)+8\DD{z}{1} C_A  \nonumber\\ 
&-4 C_A \DD{z}{0}\logs\Bigg]
+\Bigg[\delta(1-z)\Bigg(\ff{11}{108}n_f^2\pi^2
\nonumber\\  &
+ C_A^2\bigg(\ff{607}{81}
+\ff{517}{108}\pi^2-\ff{\pi^4}{80}-\ff{407}{36}
  \zeta_3\bigg) + C_A n_f \bigg(\ff{37}{18}\zeta_3
\nonumber\\  &
-\ff{11}{8}\pi^2
-\ff{82}{81}\bigg)
+ \logs \Bigg(C_F n_f + C_A
n_f\left(\ff{4}{3}
+\ff{2}{9}\pi^2\right)
\nonumber\\  &
- C_A^2\bigg(\ff{16}{3}
+\ff{11}{9}\pi^2+38\zeta_3\bigg)\Bigg) +
\logsq \Bigg(\beta_0^2
\nonumber\\ &
- C_A^2\ff{4}{3}\pi^2\Bigg)\Bigg)
+\DD{z}{0}\Bigg(C_A
n_f\left(\ff{56}{27}-\ff{4}{9}\pi^2\right)
\nonumber\\  &
+ C_A^2 \bigg(78\zeta_3 -\ff{404}{27}+\ff{22}{9}\pi^2\bigg)
+\logs\Bigg(-\ff{134}{9}C_A^2
\nonumber\\  &
+\ff{20}{9}C_A n_f + 
\ff{10}{3}C_A^2 \pi^2\Bigg)
+6\beta_0 \logsq\Bigg)
\nonumber\\  &
+ \DD{z}{1}\Bigg(-\ff{40}{9}C_A n_f 
+ C_A^2\left(\ff{268}{9}-\ff{20}{3}\pi^2\right)
\nonumber\\  &
-8 C_A\beta_0 \logs + 16
C_A^2 \logsq\Bigg)
-8 \bigg(C_A\beta_0
\nonumber\\  &
+6 C_A^2 \logs\bigg)
\DD{z}{2} + 32C_A^2 \DD{z}{3} \Bigg] + \mathcal{O}(\ep)\Bigg\}.
\end{align}
For the sake of clarity, we also retained the full scale 
dependence of the result.

\bibliographystyle{JHEP}
\bibliography{higgs_interference}

\end{document}